\documentstyle[aps,preprint]{revtex}
\begin{document}

\begin{center}
{\huge On the Charged Bose Liquid }\\ [5mm]
{\large A.A. Shanenko} \\[3mm]
{\it Bogoliubov Laboratory of Theoretical Physics \\
Joint Institute for Nuclear Research \\
141980, Dubna, Moscow region, Russia}
\end{center}

\vspace{0.3cm}

\begin{abstract}
A new approach developed in Ref.1 for investigating the spatial
boson correlations at low temperatures is considered in the particular
case of the Coulomb interaction potential. This approach is based on
integro--differential equations for the radial distribution function
and can be used beyond the weak coupling regime.
\end{abstract}
\vspace{0.5cm}

\vspace{1cm}
Description of the charged Bose liquid, or, in other words, of the
strongly correlated many--boson system with the Coulomb interparticle
potential, is a long--standing problem of the condensed matter
physics. Recently, interest in this problem has been renewed in the
connection with the noteworthy results of
applying the Bose and Bose--Fermi
liquid models~\cite{alex} to the explanation of various phenomena
inherent in the superconductors with high critical temperatures.
These models are based on the Bogoliubov approach~\cite{bogol}
implying the weak coupling regime. For the system of charged
bosons the weak coupling approximation can be used at small
values of the Brueckner parameter $r_S \sim 0.1\,.$  However,
questions requiring detailed investigations often concern
the materials with relatively low densities of charge
carriers which correspond to $r_S-$values even larger than those of
the electron gas at the metallic densities. For example, there are
the results~\cite{shan} testifying to the possibility of forming
the so--called large bipolaron (e.g. the two--electron cluster with
a large radius of electron--electron separation, see papers in
Ref.~\cite{smon}) inside the polaron environment
at the electron
densities $\sim 10^{19}\;cm^{-3}$ close to the carrier densities
of the heavily doped degenerate polar semiconductors. To investigate
the possible phase transition of the charge carrier system from the
polaron state to the bipolaron phase at these densities, one should
operate with the charged Bose liquid of large bipolarons for
$r_S \gg 1\,.$  Indeed, the ''large--bipolaron`` Bohr radius can
be expressed~\cite{shan} as
\begin{eqnarray}
a_{bip}={\hbar^2 \over m_{bip}\,Q_{bip}^2} \; \;
\left(Q_{bip}={2\,e \over \varepsilon_0^{1/2}}\right)
\nonumber\end{eqnarray}
where $\varepsilon_0$ is the static dielectric constant of the polar
material considered,$\;m_{bip}$ denotes the bipolaron mass, and
$\;e$ is the electron charge. To connect $a_{bip}$  with the canonical
electron Bohr radius $\displaystyle a_e={\hbar^2 \over m_e\,e^2}=
5.29\cdot10^{-9}\;cm\;,$ we use the reasonable estimate~\cite{smon}
$m_{bip} \approx 2\,m_{pol}~$ ($m_{pol}$ is the polaron mass) and the
relation $\displaystyle m_{pol} \sim { 16\, \alpha^4 \over 81\,\pi^2}
\;m_e^*\;$ ($m_e^*$ is the band mass) correct at the values of
the Fr\"ohlich electron--phonon coupling constant $\alpha \sim 7$
typical for the large bipolaron~\cite{smon}. This leads to
$ \displaystyle a_{bip} \sim 0.0026\,\varepsilon_0\,{\hbar^2\over
m_e^* \,e^2} \;.$ Further, with the tables of the physical properties
of polar materials~\cite{karth} we find that $\varepsilon_0 \sim 10
\div 100\,$and$\;m_e^*\sim0.1 \div 0.5\,m_e$ represent the most
frequent case. Choosing, say, $\varepsilon_0=100$ and
$m_e^*=0.1\,m_e\;$, we obtain $a_{bip} \sim 2.6 a_e =1.375
\cdot 10^{-8}\;cm\,.$
With this choice the ''large--bipolaron`` Brueckner parameter
$\displaystyle r_S=\frac{1}{a_{bip}}\left(\frac{3}{4\pi\,n}
\right)^{1/3}$ has values $\sim 20$ at the bipolaron densities $n
\sim 10^{19}\;cm^{-3}$, which proves the statement mentioned above.
Note that another choice gives even larger values of $r_S\;.$
Thus, to make the final conclusions on the possibility of the phase
transition from the Fermi state to the Bose one in the charge carrier
systems of various materials and to clear up the status of the Bose
liquid models~\cite{alex}, one should deal with the Bose
many--particle systems beyond the weak coupling regime.

The aim of this letter is to consider potentialities of a new
approach to investigating the spatial correlations of cold
bosons in the particular case of the Coulomb interaction potential.
The approach is based on integro--differential equations for the
function of the radial boson distribution that have been derived
in paper~\cite{shan1} in a way analogous to that of Ref.~\cite{shan2}.
For the charged bosons the simplest of these equations is written as
\begin{eqnarray}
\frac{\hbar^2}{m g^{1/2}(r)}\triangle g^{1/2}(r)&=&\frac{Q^2}{r}+
n \int\;\frac{Q^2}{\mid \vec r - \vec y \mid}
               \Bigl(g(y)-1\Bigr)d\vec y.
\label{eq1}\end{eqnarray}
where $g(r)$ denotes the radial distribution function, $m$ and $Q$
are the boson mass and charge, $n$ is the particle density. Besides,
$\triangle$ stands for the Laplacian. To be convinced of
the capacity of the equation (\ref{eq1}) for giving reasonable
estimates of the radial distribution function of the charged bosons,
let us consider the weak coupling regime characterized by
\begin{eqnarray}
g(r)=1-\varepsilon(r)\, , \;\;\varepsilon(r) \ll 1\,.
\label{eq2}\end{eqnarray}
Using~(\ref{eq2}) and only keeping the terms linear in
$\varepsilon(r)\,$, we arrive at the linear version of
the equation~(\ref{eq1}):
\begin{equation}
-\frac{\hbar^2}{2m}\;\triangle \varepsilon(r)={Q^2 \over r}
-n \int\; {Q^2 \over \mid \vec r - \vec y \mid}
              \; \varepsilon(y)d\vec y
\label{eq3}\end{equation}
which can easily be solved via the Fourier transformation. The
Fourier transform of $\varepsilon(r)\,$ obeys the relation
\begin{equation}
{\hbar^2\,q^2 \over 2\,m} \;\widetilde\varepsilon(q)=
{4 \pi Q^2 \over q^2} - n\,{4 \pi Q^2 \over q^2}\,
\widetilde\varepsilon(q)\, .
\label{eq4}\end{equation}
Introducing the new variable $\displaystyle x \equiv q/A$ with
$\displaystyle A^4 = 16 \pi n m Q^2 / \hbar^2\;,$ from~(\ref{eq4})
we find
\begin{equation}
\widetilde\varepsilon(q)={f_{est}(x) \over n}\, ,\;\;
f_{est}(x) = 1 - {x^4 \over x^4 + 0.5}\;.
\label{eq5}\end{equation}
It is now interesting to compare the estimate~(\ref{eq5}) with the
result~\cite{isih} of the random phase approximation (RPA)
\begin{equation}
\widetilde\varepsilon(q)={f_{RPA}(x) \over n}\, ,\;\;
f_{RPA}(x) = 1 - {x^2 \over \sqrt{ x^4 + 1}}\;.
\label{eq6}\end{equation}
Short--range correlations are determined by the behaviour of
$\widetilde\varepsilon(q)$ at $x=q/A \gg 1$ for which we have
$$ f_{est}(x) \simeq {1 \over 2\,x^4}\, , \;\;
f_{RPA}(x) \simeq {1 \over 2\,x^4} \qquad (x \gg 1).$$
There is also good agreement between $f_{est}(x)$ and $f_{RPA}(x)$
in the region $x \sim 1$ ruling the spatial correlations at the
intermediate distances close to the screening radius $1/A$:
$$f_{est}(1.5)=0.090\, , \;\;f_{RPA}(1.5)=0.086\, ;$$
$$f_{est}(1)=0.333\, , \;\;f_{RPA}(0.5)=0.293\, ;$$
$$f_{est}(0.5)=0.889\, , \;\;f_{RPA}(0.5)=0.758\, ;$$
When $x \ll 1$ (long--range correlations), we again obtain close
numerical results
$$f_{est}(0.2)=0.997\, , \;\;f_{RPA}(0.2)=0.960\, ,$$
however, $f_{est}(x)$ and $f_{RPA}(x)$ tend to unity in completely
different ways:
$$f_{est}(x) \simeq 1 - 2\,x^4 \, , \;
                       f_{RPA}(x) \simeq 1-x^2\;\;(x \ll 1)\,.$$
The comparative consideration of $f_{est}(x)$ and $f_{RPA}(x)$
allows one to make the following two conclusions. The first,
negative, is that the model presented is not able to yield
the correct behaviour of the structure factor $S(k)=1-n\,
\widetilde\varepsilon(k)$ at small $k$ and, therefore, the correct
spectrum of the low--lying excitations given by
$k^2/S(k)$(see Ref.~\cite{feyn})~. The second conclusion,
positive, consists in that the equation~(\ref{eq1}) can provide a
good estimate of the pair correlation function $h(r)=g(r)-1$ for
distances $r \ll 1/A$ and $r \sim 1/A$ exactly where the spatial
boson correlations are significant.

These conclusions have been found in the weak coupling
regime. And what can we expect, as to the results of (\ref{eq1}),
beyond the weak coupling, for $r_S > 1$. It is known~\cite{singwi}
that RPA should be corrected for $r_S > 1$ to avoid the trouble of
negative values of $g(r)$ at small particle separations. This occurs
because the short--range correlations are taken into account not in
the best way within RPA. At $r_S$ small enough the short--range
correlations do not yield any essential contribution to the
thermodynamic quantities. Therefore, this shortcoming of RPA does
not practically affect the results for $r_S < 1\,.$
On the contrary, at large $r_S$ the short--range correlations become
significant and influence the thermodynamic quantities considerably.
In this case RPA leads not only to the negative values of $g(r)$
at small $r$ but overestimates the absolute value of the ground
state energy. As to the model based on the equation~(\ref{eq1}), it
has no problem like this. Indeed, at small $r$ where
$$
\frac{Q^2}{r} \gg n \int\;\frac{Q^2}{\mid \vec r - \vec y \mid}
                                 \Bigl(g(y)-1\Bigr)d\vec y\; , $$
the expression~(\ref{eq1}) can be written as
\begin{equation}
-\frac{\hbar^2}{m}\triangle g^{1/2}(r)+
                   \frac{Q^2}{r}\; g^{1/2}(r)=0 \;.
\label{eq7}\end{equation}
It is nothing else but the Schr\"odinger equation like those
which are often used to investigate the spatial boson correlations
at small particle separations as well as in dilute
systems~\cite{shan1}. Thus, we may expect that the
relation~(\ref{eq1}) will also give reasonable results on the charged
boson correlations for $r_S > 1$.

If we have a reliable estimate of $g(r)$, we can obtain adequate
estimates of the thermodynamic quantities of the system considered.
In particular, it is possible to calculate $E\;,$ the ground state
energy of the charged bosons in the neutralizing background, with
the following well--known relation~\cite{kittel}:
\begin{equation}
\frac{E}{N}=\frac{E_{id}}{N} + \frac{1}{2}\,n
\int\limits_{0}^{1}\,d\gamma \int\,
              \Bigl(g(r,\gamma)-1\Bigr)\,\frac{Q^2}{r}\,d\vec r
\label{eq8}\end{equation}
Here $N$ denotes the number of bosons, $E_{id}$ is the energy of
noninteracting particles (in our case $E_{id}=0$~).
The quantity $\;\gamma$ is the coupling constant and $g(r,\gamma)$
denotes the solution of the equation obtained with the replacement
$Q^2 \rightarrow \gamma \, Q^2$ in~(\ref{eq1}). On the basis of
the relation~(\ref{eq5}) we find
\begin{equation}
g(r,\gamma)\,=\,1\,-\,\frac{1}{(2\pi)^3\,n}
    \int\,\left(1-\frac{q^4}{q^4 + \gamma\,0.5\,A^4}\right)\;
		      \exp(\imath\,\vec q \, \vec r\,) d\vec q
\label{eq9}\end{equation}
which can be rewritten as
\begin{equation}
g(r,\gamma)=1-{ \gamma^{1/2}\,A^2 \over 2\,\pi^2\,n\,r}\;
\int\limits_0^{\infty}\,y\,\sin(\gamma^{1/4}\,A\,r\,y)\,
\left(1-{y^4\over y^4+0.5}\right)\;dy.
\label{eq10}\end{equation}
Substituting (\ref{eq10}) into (\ref{eq8}) we obtain
\begin{equation}
\frac{E}{N}=-\frac{Q^2\,A}{\pi}\,\int\limits_0^1\;d\gamma
\int\limits_0^{\infty}\;dl\;\int\limits_0^{\infty}\;dy\;
    \;\gamma^{1/4}\;y\,\sin(l\,y)\,
                     \left(1-{y^4\over y^4+0.5}\right)\;.
\label{eq11}\end{equation}
The result of numerical integration
$$ \int\limits_0^{\infty}\;dl\;\int\limits_0^{\infty}\;dy\;
     \;y\,\sin(l\,y)\,
             \left(1-{y^4\over y^4+0.5}\right)\,=\,0.9339$$
makes it possible to derive from (\ref{eq11}) the relation
\begin{equation}
\frac{E}{N}=-0.2378\;Q^2\,A\,=\,- 0.885\;r_S^{-3/4}\;Ry
\label{eq12}\end{equation}
where $r_S$ and $Ry$ obey the standard definitions
$$r_S={m\,Q^2 \over \hbar^2}\;\left(\frac{3}{4\pi\,n}\right)^{1/3}\!,
\qquad Ry={m\,Q^4\over 2\,\hbar^2}\;.$$
The RPA result for $E/N$ can easily be found with the
relation~(\ref{eq6}). This yields the expression
\begin{equation}
\frac{E}{N}=-\frac{Q^2\,A}{\pi}\,\int\limits_0^1\;d\gamma
\int\limits_0^{\infty}\;dl\;\int\limits_0^{\infty}\;dy\;
    \; \gamma^{1/4}\;y\,\sin(l\,y)\,
                     \left(1-{y^2\over \sqrt{y^4+1}}\right)\;.
\label{eq13}\end{equation}
Numerical integration
$$ \int\limits_0^{\infty}\;dl\;\int\limits_0^{\infty}\;dy\;
     \;y\,\sin(l\,y)\,
             \left(1-{y^2\over \sqrt{y^4+1}}\right)\,=\,0.8468$$
allows one to obtain
\begin{equation}
\frac{E}{N}=-0.2156\;Q^2\,A\,=\,- 0.803\;r_S^{-3/4}\;Ry\;.
\label{eq14}\end{equation}
Thus, agreement between the results of (\ref{eq12}) and (\ref{eq14})
is within the accuracy of $10\% \,.$  Besides, it is also interesting
to make a remark on the next--to--leading contribution to the mean
energy of the system. Within the approach considered this contribution
can be calculated substituting
$$ g(r)=1-\varepsilon_1(r)-\varepsilon_2(r)
 \quad \Bigl(\varepsilon_2(r) \ll \varepsilon_1(r) \ll 1\,\Bigr) $$
into the equation (\ref{eq1}).
Omitting details of the calculations, let us only write the final
result
\begin{equation}
\frac{E}{N}=\,\Bigl(- 0.885\;r_S^{-3/4}\,+\,const\Bigr)\;Ry\,,
\quad const \approx 0.08\;
\label{eq15}\end{equation}
which can be compared with the relation
\begin{equation}
\frac{E}{N}=\,\Bigl(- 0.803\;r_S^{-3/4}\,+\,0.212\Bigr)\;Ry
\label{eq16}\end{equation}
found within the Bogoliubov scheme~\cite{bogol}.

Summarizing, let us recall the basic points of the letter once more.
The new approach of investigating the spatial correlations of
cold bosons has been considered in the particular situation of a
many--boson system with the Coulomb interaction potential.
In this case the radial distribution function $g(r)$ can be
estimated with the integro--differential equation (\ref{eq1}).
As it has been shown, this equation is able to yield reasonable
results on the short--  and intermediate--range correlations.
However, the obtained decay of the function $h(r)=g(r)-1$ at
distances essentially larger than the screening radius is not
correct. Nevertheless, Eq.(\ref{eq1}) has provided a
good estimate of the ground state energy at small $r_S\,.$ This
result gives an optimistic view on the possibility of using
Eq.(\ref{eq1}) to explore the thermodynamic behavior of the
charged Bose liquid (~$r_S > 1$~) where the influence of the
short--range correlations on the system properties is essentially
more significant than that for $r_S < 1\,.$
\\[5mm]
{\it Acknowledgement:} The author thanks A. Yu. Cherny for interest
in the subject of this letter and stimulating discussions.


\begin{thebibliography}{99}
\bibitem{shan1} A.~A.~Shanenko, Phys. Lett. {A}, 1997 (to be
published), see {\it cond-mat/9612195}.
\bibitem{alex}
R.~Friedberg, T.~D.~Lee, Phys. Lett. {\bf A 138}, 423 (1989);
R.~Micnas, J.~Ranninger, and S.~Robaszkiewicz,
Rev. Mod. Phys. {\bf 62}, 113 (1990);
A.~S.~Alexandrov, JETP Lett. {\bf 55}, 195 (1992);
A.~S.~Alexandrov, N.~F.~Mott, Supercond. Sci. Technol. {\bf 6},
215 (1993);
N.~F.~Mott, Physica {\bf C 205}, 191 (1993);
G.~Iadonisi, V.~Cataudella, D.~Ninno, M.~I.~Chiofalo, Phys. Lett.
{\bf A 196}, 359 (1995).
\bibitem{bogol} N.~N.~Bogoliubov, J. Phys. USSR {\bf 11}, 23
(1947);
L.~L.~Foldy, Phys. Rev. {\bf 124}, 649 (1961);
A.~S.~Alexandrov, W.~H.~Beere, Phys. Rev.
{\bf B51}, 5887 (1995).
\bibitem{shan} A. A. Shanenko, M.A Smondyrev, and J. T. Devreese,
Solid State Comm. {\bf 98}, 1091 (1996).
\bibitem{smon} J.~Adamowski, Phys. Rev. B {\bf 39}, 3649 (1989);
G.~Verbist, F.~M.~Peeters, and J.~T.~Devreese, Phys. Rev. B {\bf 43},
2712 (1991); F.~Bassani, M.~Geddo, G.~Iadonisi, and D.~Ninno,
Phys. Rev. B {\bf 43}, 5296 (1991); M. A. Smondyrev, J. T. Devreese,
and F. M. Peeters, Phys. Rev. B {\bf 51}, 15008 (1995).
\bibitem{karth} J.~W.~Hodby. In: `Polarons in Ionic Crystals and
Semiconductors', ed. J. T. Devreese (North-Holland, Amsterdam,
1972) p. 389; E. Kartheuser, ibid, p.717.
\bibitem{shan2} A.~A.~Shanenko, Phys. Rev. {\bf E54}, 4420 (1996).
\bibitem{isih} A. Isihara, M. Wadati, Phys. Rev. {\bf 183}, 312
(1969).
\bibitem{feyn} R. P. Feynman, Phys. Rev. {\bf 94}, 262 (1954).
\bibitem{singwi} K. S. Singwi, M. P. Tosi, R. H. Land, A. Sj\"olander,
Phys. Rev. {\bf 176}, 589 (1968); K.~S.~Singwi, A.~ Sj\"olander,
M.~ P.~Tosi, R.~H.~Land, Phys. Rev. {\bf B1}, 1044 (1970);
P.~Vashishta, K.~S.~Singwi, Phys. Rev. {\bf B6}, 875 (1972).
\bibitem{kittel} Kittel C., Quantum Theory of Solids
(Wiley, New York, 1963).

\end{thebibliography}
\end{document}